\documentclass[a4paper,11pt]{article}
\usepackage{graphicx}
\usepackage[utf8]{inputenc}
\usepackage{amsfonts}
\usepackage{amssymb}
\usepackage{amsmath}
\usepackage{dsfont}
\usepackage{bbm}
\usepackage{framed}
\usepackage{fleqn}
\usepackage[colorlinks,pdfpagelabels,pdfstartview = FitH,bookmarksopen = true,bookmarksnumbered = true,linkcolor = black,plainpages = false,hypertexnames = false,citecolor = black]{hyperref}
\usepackage{longtable}
\usepackage{pdflscape}
\usepackage[ngerman,british]{babel}
\usepackage[tight,nice]{units}
\usepackage[width=16.5cm]{geometry}
\usepackage{color,cancel}
\usepackage[multiple]{footmisc} % commas between two footnotes in the same place
\usepackage[normalem]{ulem} % provides \sout{} to cross out; useful for comments
\usepackage{url}

\definecolor{purple}{rgb}{0.5,0,0.5}

\definecolor{darkgreen}{rgb}{0.2,0.8,0.2}
\newcommand{\Tabref}[1]{Table~\ref{#1}}

\newcommand{\E}[1]{\ensuremath{\mathrm{E}_{#1}}} % e.g. \E{8}

\newcommand{\SO}[1]{\ensuremath{\mathrm{SO}(#1)}}
\newcommand{\SU}[1]{\ensuremath{\mathrm{SU}(#1)}}
\newcommand{\U}[1]{\ensuremath{\mathrm{U}(#1)}}
\newcommand{\Z}[1]{\ensuremath{\mathbbm{Z}_{#1}}} % z_N ->\Z{N}

\newcommand{\rep}[1]{\ensuremath\boldsymbol{#1}}

\numberwithin{equation}{section}
\numberwithin{table}{section}

\begin{document}

\begin{titlepage}

\vspace*{-3.0cm}
\begin{flushright}
\normalsize{LMU-ASC 05/14}\\
\normalsize{TUM-HEP 932/14}
\end{flushright}

\vspace*{1.0cm}

\begin{center}
{\Large\textbf{Geography of Fields in Extra Dimensions:\\String Theory Lessons for Particle Physics}}

\vspace{1cm}

\textbf{
Hans~Peter~Nilles$^{a}$ and Patrick~K.S.~Vaudrevange$^{b}$
}
\\[5mm]
$^a$~Bethe Center for Theoretical Physics\\
and\\
Physikalisches Institut der Universit\"at Bonn,\\
Nussallee 12, 53115 Bonn, Germany\\
\vspace{2mm}
$^b$~Excellence Cluster Universe, Technische Universit\"at M\"unchen,\\
Boltzmannstr. 2, D-85748, Garching, Germany\\
and\\
Arnold Sommerfeld Center for Theoretical Physics, LMU\\
Theresienstra\ss e 37, 80333 M\"unchen, Germany
\end{center}

\vspace{1cm}

\vspace*{1.0cm}

\begin{abstract}
String theoretical ideas might be relevant for particle physics model 
building. Ideally one
would hope to find a unified theory of all fundamental interactions. There 
are only few
consistent string theories in $D=10$ or $11$ space-time dimensions, but a 
huge landscape in
$D=4$. We have to explore this landscape to identify models that describe 
the known
phenomena of particle physics. Properties of compactified six spatial 
dimensions are
crucial in that respect. We postulate  some useful rules to investigate 
this landscape
and construct realistic models. We identify common properties of the 
successful models and 
formulate lessons for further model building.
\end{abstract}

\end{titlepage}

\newpage

%%%%%%%%%%%%%%%%%%%%%%%%%%%%%%%%%%%%%%%%%%%%%%%%%%%%%%%%%%%%%%%%%%%%%%%%%%%%%%%%%%%%%%
\section{Introduction}
%%%%%%%%%%%%%%%%%%%%%%%%%%%%%%%%%%%%%%%%%%%%%%%%%%%%%%%%%%%%%%%%%%%%%%%%%%%%%%%%%%%%%%

One of the main goals of string theory is the inclusion of the Standard Model (SM) 
of particle physics in an ultraviolet complete and consistent theory of quantum gravity.
The hope is a unified theory of all fundamental interactions: gravity as well as strong
and electroweak interactions within the $\SU{3}\times\SU{2}\times\U{1}$ SM. Recent support
for the validity of the particle physics Standard Model is the 2012 discovery of the 
``so-called'' Higgs boson. 

How does this fit into known string theory? Ideally one would have hoped to derive the
Standard Model from string theory itself, but up to now such a program has not (yet) been
successful. It does not seem that the SM appears as a prediction of string theory. In
view of that we have to ask the question whether the SM can be embedded in string
theory. If this is possible we could then scan the successful models and check specific
properties that might originate from the nature of the underlying string theory. 

Known superstring theories are formulated in $D=10$ space time dimensions (or
$D=11$ for M theory) while the SM describes physics in $D=4$. The connection
between $D=10$ and $D=4$ requires the compactification of six spatial dimensions. 
The rather few superstring theories in
$D=10$ give rise to a plethora of theories in $D=4$ with a wide spectrum of
properties. The search for the SM and thus the field of so-called ``String
Phenomenology'' boils down to a question of exploring this compactification process
in detail.

But how should we proceed? As the top-down approach is not successful we should 
therefore analyse in detail the properties of the SM and then use a bottom-up approach 
to identify those regions of the ``string landscape'' where the SM is most likely to reside. 
This will provide us with a set of ``rules'' for $D=4$ model constructions of string theory 
towards the constructions of models that resemble the SM.

The application of these rules will lead us to ``fertile patches'' of the string landscape
with many explicit candidate models. Given these models we can then try to identify
those properties of the models that make them successful. They teach us some
lessons towards selecting the string theory in $D=10$ as well as details of the process
of compactification.

In the present paper we shall describe this approach to ``string phenomenology''. In 
section~\ref{sec:fivegoldenrules} we shall start with ``five golden rules'' as they have been 
formulated some time ago~\cite{Nilles:2004ej}. These rules have been derived in a bottom-up 
approach exploiting the particular properties of quark- and lepton representations in the SM. 
They lead to some kind of (grand) unified picture favouring $\SU{5}$ and $\SO{10}$ symmetries 
in the ultraviolet. However, these rules are not to be understood as strict rules for string model 
building. You might violate them and still obtain some reasonable models. But, as we shall see, 
life is more easy if one follows these rules.

In section~\ref{sec:minilandscape} we shall start explicit model building along these lines. We 
will select one of those string theories that allow for an easy incorporation of the rules within 
explicit solvable compactifications. This leads us to orbifold compactifications of the heterotic
$\E{8}\times\E{8}$ string theory~\cite{Dixon:1985jw,Dixon:1986jc} as an example. We shall 
consider this example in detail and comment on generalizations and alternatives later. The 
search for realistic models starts with the analysis of the so-called $\Z{6}$-II 
orbifold~\cite{Kobayashi:2004ya,Buchmuller:2004hv,Buchmuller:2005jr,Buchmuller:2006ik,Lebedev:2006kn,Lebedev:2007hv,Ratz:2007my,Lebedev:2008un}. 
We define the search strategy in detail and present the results known as the 
``MiniLandscape''~\cite{Lebedev:2006kn, Lebedev:2008un}, a fertile patch of the string landscape 
for realistic model building. We analyse the several hundred models of the MiniLandscape towards 
specific properties, as e.g.\ the location of fields in extra-dimensional space. The emerging 
picture leads to a notion of ``Local Grand Unification'', where some of the more problematic 
aspects of grand unification (GUT) can be avoided. We identify common properties of the successful 
models and formulate ``lessons'' from the MiniLandscape that should join the ``rules'' for realistic 
model building.

Section~\ref{sec:orbifoldlandscape} will be devoted to the construction of new, explicit MSSM-like 
models using all $\Z{N}$ and certain $\Z{N}\times\Z{M}$ orbifold geometries resulting in approximately 
12000 orbifold models. Then, in section~\ref{sec:rulesorbifoldlandscape} we shall see how the lessons 
of the MiniLandscape will be tested in this more general ``OrbifoldLandscape''.

In section~\ref{sec:generallandscape} we shall discuss alternatives to orbifold compactifications, as 
well as model building outside the heterotic $\E{8}\times\E{8}$ string. The aim is a unified picture 
of rules and lessons for successful string model building. Section~\ref{sec:nvconclusions} will be 
devoted to conclusions and outlook.

%%%%%%%%%%%%%%%%%%%%%%%%%%%%%%%%%%%%%%%%%%%%%%%%%%%%%%%%%%%%%%%%%%%%%%%%%%%%%%%%%%%%%%
\section{Five golden rules}
%%%%%%%%%%%%%%%%%%%%%%%%%%%%%%%%%%%%%%%%%%%%%%%%%%%%%%%%%%%%%%%%%%%%%%%%%%%%%%%%%%%%%%
\label{sec:fivegoldenrules}

Let us start with a review of the ``Five golden rules for superstring phenomenology'', 
which can be seen as phenomenologically motivated guidelines to successful string 
model building~\cite{Nilles:2004ej}. The rules can be summarized as follows: we need
\begin{enumerate}
\item spinors of $\SO{10}$ for SM matter
\item incomplete GUT multiplets for the Higgs pair
\item repetition of families from geometrical properties of the compactification space
\item $\mathcal{N} = 1$ supersymmetry
\item $R$-parity and other discrete symmetries
\end{enumerate}
Let us explain the motivation for these rules in some detail in the following.

\subsection{Rule I: Spinors of $\boldsymbol{\SO{10}}$ for SM Matter}
\label{sec:firstgoldenrules}

It is a remarkable fact that the spinor $\rep{16}$ of $\SO{10}$ is the unique irreducible 
representation that can incorporate exactly one complete generation of quarks and leptons, 
including the right-handed neutrino. Thereby, it can explain the absence of 
gauge-anomalies in the Standard Model for each generation separately. Furthermore, 
it offers a simple explanation for the observed ratios of the electric charges of all 
elementary particles. In addition, there is a strong theoretical motivation for Grand 
Unified Theories like $\SO{10}$ from gauge coupling unification at the GUT scale $M_\text{GUT} 
\approx 3 \times 10^{16}$ GeV. Hence, the first golden rule for superstring phenomenology 
suggests to construct string models in such a way that at least some generations of quarks 
and leptons reside at a location in compact space, where they are subject to a larger gauge 
group, like $\SO{10}$. Hence, these generations come as complete representations of that 
larger group, e.g.\ as $\rep{16}$ of $\SO{10}$.

The heterotic string offers this possibility through the natural presence of the exceptional 
Lie group $\E{8}$, which includes an $\SO{10}$ subgroup and its spinor representation. 
Furthermore, using orbifold compactification the four-dimensional Standard Model gauge group
can be enhanced to a local GUT, i.e\ to a GUT group like $\SO{10}$ which is realized 
locally at an orbifold singularity in extra dimensions. In addition, there are matter fields 
(originating from the so-called twisted sectors of the orbifold) localised at these special 
points in extra dimensions and hence they appear as complete multiplets of the local GUT group, for 
example as $\rep{16}$-plets of $\SO{10}$.

On the other hand, the spinor of $\SO{10}$ is absent in (perturbative) type II string theories, 
which can be seen as a drawback of these theories. Often this drawback manifests itself in an 
unwanted suppression of the top quark Yukawa coupling. On the other hand, F-theory 
(and M-theory) can cure 
this through the non-perturbative construction of exceptional Lie groups like e.g. $\E{6}$. 
When two seven-branes with $\SO{10}$ gauge group intersect in the extra dimensions, a local 
GUT can appear at the intersection. There, the gauge group can be enhanced to a local $\E{6}$ 
and a spinor of $\SO{10}$ can appear as matter representation.

\subsection{Rule II: Incomplete GUT Multiplets for the Higgs Pair}

Beside complete spinor representations of $\SO{10}$ for quarks and leptons, the (supersymmetric 
extension of the) Standard Model needs split, i.e.\ incomplete $\SO{10}$ multiplets for the gauge 
bosons and the Higgs(--pair). Their unwanted components inside a full GUT multiplet would induce 
rapid proton decay and hence need to be ultra-heavy. In the case of the Higgs doublet, this 
problem is called the doublet--triplet splitting problem, because for the smallest GUT $\SU{5}$ 
a Higgs field would reside in a five-dimensional representation of $\SU{5}$, which includes 
beside the Higgs doublet an unwanted Higgs triplet of $\SU{3}$. This problem might determine 
the localisations of the Higgs pair and of the gauge bosons in the compactification space: 
they need to reside at a place in extra dimensions where they feel the breaking of the 
higher-dimensional GUT to the 4D SM gauge group. Hence,  incomplete GUT multiplets, e.g.\ for 
the Higgs, can appear. This is the content of the second golden rule. 

In this way local GUTs exhibit grand unified gauge symmetries only at some special ``local'' 
surroundings in extra dimensions, while in 4D the GUT group seems to be broken down to the 
Standard Model gauge group. This allows us to profit from some of the nice properties of 
GUTs (like complete representations for matter as described in the first golden rule), while 
avoiding the problematic properties (like doublet--triplet splitting).

In the case of the heterotic string on orbifolds the so-called untwisted sector (i.e.\ the 
10D bulk) can naturally provide such split $\SO{10}$ multiplets for the gauge bosons and the 
Higgs. In particular, when the orbifold twist acts as a $\Z{2}$ in one of the three complex 
extra dimensions, one can obtain an untwisted Higgs pair that is vector-like with respect 
to the full (i.e.\ observable and hidden) gauge group. Combined with an (approximate) 
$R$-symmetry this can yield a solution to the $\mu$-problem of the MSSM. Furthermore, as all 
charged bulk fields originate from the 10D $\E{8}\times\E{8}$ vector multiplet this scenario 
naturally yields gauge--Higgs--unification.

Finally, an untwisted Higgs pair in the framework of heterotic orbifolds can relate the top 
quark Yukawa coupling to the gauge coupling and hence give a nice explanation for the large 
difference between the masses of the third generation compared to the first and second one. 
In order to achieve this, the top quark needs to originate either from the bulk (as it is often 
the case in the MiniLandscape~\cite{Lebedev:2006kn}  of $\Z{6}$-II orbifolds) or from an 
appropriate fixed torus, i.e.\ a complex codimension one singularity in the extra dimensions.

\subsection{Rule III: Repetition of Families}

The triple repetition of quarks and leptons as three generations with the same gauge 
interactions but different masses is a curiosity within the Standard Model and asks for 
a deeper understanding. One approach from a bottom-up perspective is to engineer a 
so-called flavour symmetry: one introduces a (non-Abelian) symmetry group, discrete or 
gauge, and unifies the three generations of quarks and leptons in, for example, a single 
three-dimensional representation of that flavour group. However, as the Yukawa 
interactions violate the flavour symmetry, it must be broken spontaneously by the 
vacuum expectation value of some Standard Model singlet, the so-called flavon. This 
might explain the mass ratios and mixing patterns of quarks and leptons.

The third golden rule for superstring phenomenology asks for the origin of such a flavour 
symmetry. The rule suggests to choose the compactification space such that some of its 
geometrical properties lead to a repetition of families and hence yields a discrete flavour 
symmetry. In this case, the repetition of the family structure comes from topological 
properties of the compact manifold.
Within the framework of type II string theories, the number of families can be related to 
intersection numbers of D-branes in extra dimensions, while for the heterotic string 
it can be due to a degeneracy between orbifold singularities. In the latter case, one can easily 
obtain non-Abelian flavour groups which originate from the discrete symmetry transformations that 
interchange the degenerate orbifold singularities, combined with a stringy selection 
rule that is related to the orbifold space group~\cite{Kobayashi:2006wq}. In any case the
number of families will be given by geometrical and topological properties of the
compact six-dimensional manifold.

\subsection{Rule IV: $\boldsymbol{\mathcal{N} = 1}$ Supersymmetry}

Superstring theories are naturally equipped with $\mathcal{N}=1$ or 2 supersymmetry in 
10D. However, generically all supersymmetries are broken by the compactification to 4D. 
The fourth golden rule suggests to choose a ``non-generic'' compactification space such 
that $\mathcal{N}=1$ survives in 4D. Examples for such special spaces are Calabi--Yaus, 
orbifolds and orientifolds. Motivation for this is a solution of the so-called
``hierarchy problem'' between the weak scale (a TeV) and the string (Planck) scale.
Supersymmetry can stabilize this large hierarchy. Since such a supersymmetry appears
naturally in string theory, we assume that $\mathcal{N}=1$ supersymmetry will survive
down to the TeV-scale.

\subsection{Rule V: $\boldsymbol{R}$-Parity and other Discrete Symmetries}

Apart from the gauge symmetries of string theory, we need more symmetries to
describe particle physics phenomena of the supersymmetric Standard Model.
These could provide the desired textures of Yukawa couplings, explain the
absence of flavour changing neutral currents, help to avoid too fast proton 
decay, provide a stable particle for cold dark matter and solve the
so-called $\mu$-problem. We know that (continuous) global symmetries might not be
compatible with gravitational interactions. Hence, local discrete
symmetries might play this role in string theory.

One of these symmetries is the well-known matter parity of the minimal supersymmetric
extension of the Standard Model (MSSM): it forbids proton decay via dim. 4 operators 
and leads to a stable neutral WIMP candidate. Other discrete gauge symmetries are
required to explain the flavour structure of quark/lepton masses and mixings.

%%%%%%%%%%%%%%%%%%%%%%%%%%%%%%%%%%%%%%%%%%%%%%%%%%%%%%%%%%%%%%%%%%%%%%%%%%%%%%%%%%%%%%
\section{The MiniLandscape}
%%%%%%%%%%%%%%%%%%%%%%%%%%%%%%%%%%%%%%%%%%%%%%%%%%%%%%%%%%%%%%%%%%%%%%%%%%%%%%%%%%%%%%
\label{sec:minilandscape}

As we have seen in our review in section~\ref{sec:fivegoldenrules}, the five golden 
rules~\cite{Nilles:2004ej} naturally ask for exceptional Lie groups. $\SO{10}$, although it 
is not an exceptional group, fits very well in the chain of exceptional groups 
$\E{8}\rightarrow\E{7}\rightarrow\E{6}\rightarrow\SO{10}\rightarrow\SU{5}\rightarrow\text{SM}$. 
Therefore, the $\E{8}\times\E{8}$ heterotic string is the prime candidate and we choose it as 
our starting point. Alternatives to obtain $\E{8}$ in string theory are M- and F-theory, where such
gauge groups can appear in non-perturbative constructions.

\begin{figure}[h!]
\begin{center}
\resizebox{0.6\textwidth}{!}{%
\includegraphics{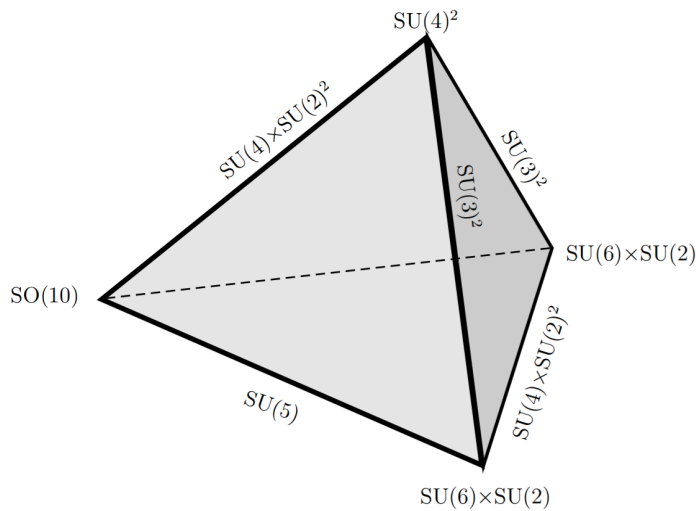}
}\end{center}
\vspace{-4mm}
\caption{Gauge group topography from Ref.~\cite{Nilles:2008gq}. At different fixed points (corners of the
tetrahedron), \E8 gets broken to different subgroups (\U1 factors are
suppressed). At the edges we display the intersection of the two local gauge
groups realised at the corners. The 4D gauge group is the standard model gauge group.}
\label{fig:gaugegrouptopo}
\end{figure}

The implementation of the rules in string theory started with the consideration of orbifold 
compactifications of the $\E{8}\times\E{8}$ heterotic string. This lead to the so-called 
``heterotic brane world''~\cite{Forste:2004ie} where toy examples have been constructed in the 
framework of the $\Z{2}\times\Z{2}$ orbifold. There, the explicit ``geographical'' properties of 
fields in extra dimensions have been presented and the local GUTs at the orbifold fixed points 
were analysed, see e.g.\ Fig.~\ref{fig:gaugegrouptopo}.

%%%%%%%%%%%%%%%%%%%%%%%%%%%%%%%%%%%%%%%%%%%%%%%%%%%%%%%%%%%%%%%%%%%%%%%%%%%%%%%%%%%%%%%%%%%%
\subsection{Exploring the $\Z{6}$-II orbifold}
%%%%%%%%%%%%%%%%%%%%%%%%%%%%%%%%%%%%%%%%%%%%%%%%%%%%%%%%%%%%%%%%%%%%%%%%%%%%%%%%%%%%%%%%%%%%
\label{sec:reviewminilandscape}

\begin{figure}[t!]
\resizebox{0.99\textwidth}{!}{%
  \includegraphics{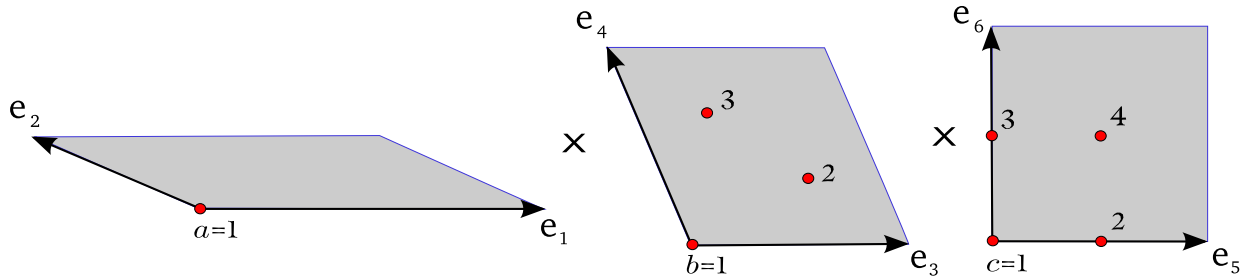}
}
\vspace{-2mm}
\caption{The six-dimensional torus $(e_1,\ldots,e_6)$ of the $\Z{6}$-II orbifold. In the $\theta$-, $\omega$-twisted sector the second, third torus is left invariant, respectively, while in the $\theta\omega$-sector there are fixed points (labelled by $a,b,c$).
}
\label{fig:thetaomegasector}
\end{figure}

A first systematic attempt at realistic model constructions~\cite{Lebedev:2006kn,Lebedev:2008un} 
was based on the $\Z{6}$-II orbifold~\cite{Kobayashi:2004ya} of the $\E{8}\times\E{8}$ heterotic 
string. This orbifold considers a six-torus defined by the six-dimensional lattice of 
$\text{G}_2\times\SU{3}\times\SO{4}$ modded out by two twists, each acting in four of the six 
extra dimensions: $\theta$ of order 2 ($\theta^2=1$) and $\omega$ of order 3 ($\omega^3=1$), see 
Fig.~\ref{fig:thetaomegasector}.

In Ref.~\cite{Lebedev:2006kn} the embedding of the twists into the $\E{8}\times\E{8}$ gauge group 
was chosen in such a way that at an intermediate step there are local $\SO{10}$ GUTs with localised 
$\rep{16}$-plets for quarks and leptons. This choice can be motivated by rule I, as discussed in the 
previous section. Further breakdown of the gauge group to $\SU{3}\times\SU{2}\times\U{1}$ is 
induced by two orbifold Wilson lines~\cite{Ibanez:1986tp}. In this set-up, a scan for realistic models 
was performed using the following strategy:

\begin{itemize}
\item choose appropriate Wilson lines (and identify inequivalent models)
\item SM gauge group $\SU{3}\times\SU{2}\times\U{1}_\text{Y} \subset\E{8}$ times a hidden sector
\item Hypercharge $\U{1}_\text{Y}$ is non-anomalous and in $\SU{5}$ GUT normalisation
\item (net) number of three generations of quarks and leptons
\item at least one Higgs pair
\item exotics are vector-like w.r.t.\ the SM gauge group and can be decoupled 
\end{itemize}

Using the above criteria, the computer assisted search led to a total of some 200 and 300 MSSM-like 
models in Refs.~\cite{Lebedev:2006kn} and~\cite{Lebedev:2008un}, respectively. The models typically 
have additional vector-like exotics as well as unbroken $\U{1}$ gauge symmetries, one of which is 
anomalous. This anomaly induces an Fayet--Iliopoulos--term (FI-term), hence a breakdown of the 
additional $\U{1}$'s and thus allows for a decoupling of the vector-like exotics. Explicit examples 
are given in Ref.~\cite{Lebedev:2007hv} as benchmark models. 

All fields of the models can be attributed to certain sectors with specific geometrical properties. 
In the present case there is an untwisted sector with fields in 10D (bulk), as well as 
twisted sectors where fields are localised at certain points (or two-tori) in the six-dimensional 
compactified space. The $\theta\omega$ twisted sector (Fig.~\ref{fig:thetaomegasector}) has fixed 
points and thus yields fields localised at these points in extra dimensions that can only propagate 
in our four-dimensional space--time. The $\theta$ and $\omega$ twisted sectors, in contrast, have 
fixed two-tori in extra dimensions. Fields in these sectors are confined to six space--time dimensions. 
Many properties of the models depend on these ``geographic'' properties of the fields in extra 
dimensions. For example, Yukawa couplings between matter and Higgs fields and in particular their 
coupling strengths are determined by the ``overlap'' of the fields in extra dimensions.

%%%%%%%%%%%%%%%%%%%%%%%%%%%%%%%%%%%%%%%%%%%%%%%%%%%%%%%%%%%%%%%%%%%%%%%%%%%%%%%%%%%%%%%%%%%%
\subsection{Lessons from the $\Z{6}$-II MiniLandscape}
%%%%%%%%%%%%%%%%%%%%%%%%%%%%%%%%%%%%%%%%%%%%%%%%%%%%%%%%%%%%%%%%%%%%%%%%%%%%%%%%%%%%%%%%%%%%

Given this large sample of realistic models, we can now analyse their properties and look for
similarities and regularities. Which geometrical and geographical properties in extra 
dimensions are important for realistic models?

By construction, all the models have observable sector gauge group $\SU{3}\times\SU{2}\times\U{1}$
and possibly some hidden sector gauge group relevant for supersymmetry breakdown. There is a 
net number of three generations of quarks and leptons and at least one pair of Higgs 
doublets $H_u$ and $H_d$. The Higgs--triplets are removed and the doublet--triplet splitting 
problem is solved. A first question concerns a possible ``$\mu$-term'': $\mu H_u H_d$ and we 
shall start our analysis with the Higgs--system, following the discussion of Ref.~\cite{Nilles:2014epj}.

%%%%%%%%%%%%%%%%%%%%%%%%%%%%%%%%%%%%%%%%%%%%%%%%%%%%%%%%%%%%%%%%%%%%%%%%%%%%%%%%%%%%%%%%%%%%%%
\subsubsection{Lesson 1: Higgs--doublets from the bulk}
%%%%%%%%%%%%%%%%%%%%%%%%%%%%%%%%%%%%%%%%%%%%%%%%%%%%%%%%%%%%%%%%%%%%%%%%%%%%%%%%%%%%%%%%%%%%%

The Higgs--system is vector-like and a $\mu$-term $\mu H_u H_d$ is potentially allowed. As this 
is a term in the superpotential we would like to understand why $\mu$ is small compared to the GUT-scale: 
This is the so-called $\mu$-problem. To avoid this problem one could invoke a symmetry that 
forbids the term. However, we know that $\mu$ has to be non-zero. Hence, the symmetry has to be broken 
and this might reintroduce the $\mu$-problem again. In string theory the problem is often amplified 
since typically we find several (say $N$) Higgs doublet pairs. In the procedure to remove the vector-like 
exotics (as described above) we have to make $N-1$ pairs heavy while keeping one light. In fact, in many 
cases the small $\mu$-parameter is the result of a specific fine-tuning in such a way to remove all doublet 
pairs except for one. We do not consider this as a satisfactory solution. Fortunately, the models of the 
MiniLandscape are generically not of this kind.

Many MiniLandscape models provide one Higgs pair that resists all attempts to remove it. This is 
related to a discrete R-symmetry~\cite{Lebedev:2007hv} that can protect the 
$\mu$-parameter in the following way: In some cases the discrete R-symmetry is enlarged to an approximate 
$\U{1}_R$~\cite{Kappl:2008ie,Brummer:2010fr}.\footnote{In addition, $\U{1}_R$ symmetries can explain 
vanishing vacuum energy in SUSY vacua.} Therefore, a $\mu$-parameter is generated at a higher 
order $M$ in the superpotential $W$, where the approximate $\U{1}_R$ is broken to its exact discrete subgroup. 
This yields a suppression $\mu\sim\langle W\rangle\sim\epsilon^M$, where $\epsilon < 1$ is set by the FI parameter. 

The crucial observation for this mechanism to work is the localisation of the Higgs pair $H_u$ and $H_d$ 
in agreement with our second golden rule: both reside in the 10D bulk originating from gauge fields 
in extra dimensions. Furthermore, the Higgs pair is vector-like with respect to all symmetries, gauge and discrete. 
This is related to the $\Z{2}$ orbifold action in one of the two-tori. Hence, each term in the superpotential 
$f(\Phi_i) \subset W$ also couples to the Higgs pair, i.e.\ $f(\Phi_i) H_u H_d \subset W$. As SUSY breakdown 
requires a non-vanishing VEV of the superpotential the $\mu$-term is related to the gravitino mass, 
i.e.\ $\mu = f(\langle \Phi_i\rangle) = \langle W\rangle \sim \epsilon^M \sim m_{3/2}$. This is a 
reminiscent of a field theoretical mechanism first discussed in Ref.~\cite{Casas:1992mk}.

%%%%%%%%%%%%%%%%%%%%%%%%%%%%%%%%%%%%%%%%%%%%%%%%%%%%%%%%%%%%%%%%%%%%%%%%%%%%%%%%%%%%%%%%%
\subsubsection{Lesson 2: Top-quark from the bulk}
%%%%%%%%%%%%%%%%%%%%%%%%%%%%%%%%%%%%%%%%%%%%%%%%%%%%%%%%%%%%%%%%%%%%%%%%%%%%%%%%%%%%%%%%%

Among all quarks and leptons the top-quark is very special: its large mass requires a large 
top-quark Yukawa coupling. Many MiniLandscape models address this naturally via the 
localisation of the top-quark in extra dimensions: both $(t, b)$ and $\bar{t}$ reside in the 
10D bulk, along with the Higgs pair. Hence, we have gauge-Yukawa unification and the 
trilinear Yukawa coupling of the top is given by the gauge coupling. 

Typically the top-quark is the only matter field with trilinear Yukawa coupling. The location 
of the other fields of the third family is strongly model-dependent, but in general they are 
distributed over various sectors: the third family could be called a ``patchwork family''.

%%%%%%%%%%%%%%%%%%%%%%%%%%%%%%%%%%%%%%%%%%%%%%%%%%%%%%%%%%%%%%%%%%%%%%%%%%%%%%%%%%%%%%%%%%
\subsubsection{Lesson 3: Flavour symmetry for the first two families}
%%%%%%%%%%%%%%%%%%%%%%%%%%%%%%%%%%%%%%%%%%%%%%%%%%%%%%%%%%%%%%%%%%%%%%%%%%%%%%%%%%%%%%%%%%

The first two families are found to be located at fixed points in extra dimensions (Fig.~\ref{fig:thetaomegasector}). 
As such they live at points of enhanced symmetries, both gauge and discrete.

The discrete symmetry is the reason for their suppressed Yukawa couplings. In the $\Z{6}$-II example 
shown in the figure two families live at adjacent fixed points in the third extra-dimensional torus: 
one family is located at $a=b=c=1$, the other at $a=b=1$ and $c=3$ (see Fig.~\ref{fig:thetaomegasector}). 
Technically, this is a consequence of a vanishing Wilson line in the $e_6$ direction. This leads to a 
$\text{D}_4$ flavour symmetry~\cite{Kobayashi:2004ya,Kobayashi:2006wq,Nilles:2012cy}. The two localised 
families form a doublet, while the third family transforms in a one-dimensional representation of $\text{D}_4$. 
This set-up forbids sizeable flavour changing neutral currents and thus relieves the so-called ``flavour 
problem''. Furthermore, the geometric reason for small Yukawa couplings of the first and second family 
is their minimal overlap with the bulk Higgs fields. This leads to Yukawa couplings of higher order 
and a hierarchical generation of masses based on the Froggatt--Nielsen mechanism~\cite{Froggatt:1978nt}, 
where the FI-term provides a small parameter $\epsilon$ that controls the pattern of masses.

In addition, the first two families live at points of enhanced gauge symmetries and therefore 
build complete representations of the local grand unified gauge group, e.g.\ as $\rep{16}$-plets 
of $\SO{10}$. Hence, they enjoy the successful properties of ``Local Grand Unification'' 
outlined in the first golden rule.

%%%%%%%%%%%%%%%%%%%%%%%%%%%%%%%%%%%%%%%%%%%%%%%%%%%%%%%%%%%%%%%%%%%%%%%%%%%%%%%%%%%%%%%%%%
\subsubsection{Lesson 4: The pattern of SUSY breakdown}
%%%%%%%%%%%%%%%%%%%%%%%%%%%%%%%%%%%%%%%%%%%%%%%%%%%%%%%%%%%%%%%%%%%%%%%%%%%%%%%%%%%%%%%%%

The question of supersymmetry breakdown is a complicated process and we shall try to extract 
some general lessons that are rather model-independent. Specifically we would consider gaugino 
condensation in the hidden sector~\cite{Nilles:1982ik,Ferrara:1982qs,Derendinger:1985kk,Dine:1985rz}
realized explicitly in the MiniLandscape~\cite{Lebedev:2006tr}, see also 
section~\ref{sec:orbifoldlandscaperule4}.

A reasonable value for the gravitino mass can be obtained if the dilaton is fixed at a realistic value 
$1/g^2(M_\text{GUT}) = \text{Re}S \approx 2$. Thus, the discussion needs the study of moduli stabilization, 
which, fortunately, we do not have to analyse here. In fact we can rely on some specific pattern of 
supersymmetry breaking which seems to be common in various string theories, first observed in the 
framework of Type IIB theory~\cite{Choi:2004sx,Choi:2005ge,Lebedev:2006qq,Vera:2012,Lebedev:2005ge,LoaizaBrito:2005fa,Lebedev:2006qc} 
and later confirmed in the heterotic case~\cite{Lowen:2008fm,Lowen:2009nr}: so-called ``mirage mediation''. 
Its source is a suppression of the tree level contribution in modulus mediation (in particular for gaugino 
masses and A-parameters). The suppression factor is given by the logarithm of the ``hierarchy'' 
$\log(M_\text{Planck}/m_{3/2})$, which numerically is of the order $4\pi^2$. Non-leading terms suppressed 
by loop factors can now compete with the tree-level contribution. In its simplest form the loop corrections 
are given by the corresponding $\beta$-functions, leading to ``anomaly mediation'' if the tree level 
contribution is absent. 
Without going into detail, let us just summarise the main properties of mirage mediation:
\begin{itemize}
\item gaugino masses and A-parameters are suppressed compared to the
      gravitino mass by the factor\ \  $\log(M_\text{Planck}/m_{3/2})$
\item we obtain a compressed pattern of gaugino masses (as the $SU(3)$
      $\beta$-function is negative while those of $SU(2)$ and $U(1)$ are positive)
\item soft scalar masses $m_0$ are more model-dependent. In general we would
      expect them to be as large as $m_{3/2}$ \cite{Lebedev:2006qq}.
\end{itemize}
The models of the MiniLandscape inherit this generic picture. But they also teach us something new on the soft 
scalar masses, which results in lesson 4. The scalars reside in various localisations in the extra dimensions 
that feel SUSY in different ways: First, the untwisted sector is obtained from simple torus compactification of the 
10D theory leading to extended $\mathcal{N}=4$ supersymmetry in $D=4$. Hence, soft terms of bulk fields are 
protected (at least at tree level) and broken by loop corrections when they communicate to sectors with less 
SUSY. Next, scalars localised on fixed tori feel a remnant $\mathcal{N}=2$ SUSY and might be protected as 
well. Finally, fields localised at fixed points feel only $\mathcal{N}=1$ SUSY and are not further 
protected~\cite{Krippendorf:2012ir,Badziak:2012yg}. Therefore, we expect soft terms $m_0\sim m_{3/2}$ for the 
localised first two families, while other (bulk) scalar fields, in particular the Higgs bosons and the stop, 
feel a protection from extended SUSY. Consequently, their soft masses are suppressed compared to $m_{3/2}$ (by 
a loop factor of order $1/4\pi^2$). This constitutes lesson 4 of the MiniLandscape.

%%%%%%%%%%%%%%%%%%%%%%%%%%%%%%%%%%%%%%%%%%%%%%%%%%%%%%%%%%%%%%%%%%%%%%%%%%%%%%%%%%%%%%
\section{The OrbifoldLandscape}
%%%%%%%%%%%%%%%%%%%%%%%%%%%%%%%%%%%%%%%%%%%%%%%%%%%%%%%%%%%%%%%%%%%%%%%%%%%%%%%%%%%%%%
\label{sec:orbifoldlandscape}

The 10D heterotic string compactified on a six-dimensional toroidal orbifolds provides 
an easy and calculable framework for string phenomenology~\cite{Dixon:1985jw,Dixon:1986jc}. 
A toroidal orbifold is constructed by a six-dimensional torus divided out by some of its 
discrete isometries, the so-called point group. For simplicity we assume this discrete symmetry 
to be Abelian. Combined with the condition on $\mathcal{N}=1$ supersymmetry in 4D one is left 
with certain $\Z{N}$ and $\Z{N}\times\Z{M}$ groups, in total 17 different choices. For each 
choice, there are in general several inequivalent possibilities, e.g.\ related to the 
underlying six-torus. Recently, these possibilities have been classified using methods from 
crystallography, resulting in 138 inequivalent orbifold geometries with Abelian point 
group~\cite{Fischer:2012qj}.

The orbifolder~\cite{Nilles:2011aj} is a powerful computer program to analyse these Abelian 
orbifold compactifications of the heterotic string. The program includes a routine to 
automatically generate a huge set of consistent (i.e.\ modular invariant and hence anomaly-free) 
orbifold models and to identify those that are phenomenologically interesting, e.g.\ that 
are MSSM-like. 

A crucial step in this routine is the identification of inequivalent orbifold models in order to 
avoid an overcounting: even though the string theory input parameters of two models (i.e.\ so-called 
shifts and orbifold Wilson lines) might look different, the models can be equivalent and share, for 
example, the same massless spectrum and couplings. The current version (1.2) of the orbifolder uses 
simply the massless spectrum in terms of the representations under the full non-Abelian gauge group 
in order to identify inequivalent models. However, there are typically five to ten \U1 factors and the 
corresponding charges are neglected for this comparison of spectra, because they are highly  
dependent on the choice of \U1 basis. As pointed out by Groot Nibbelink and Loukas~\cite{Nibbelink:2013lua} 
one can easily improve this by using in addition to the non-Abelian representations also the 
$\U{1}_\text{Y}$ hypercharge as it is uniquely defined for a given MSSM model. We included this 
criterion into the orbifolder. However, it turns out that using this refined comparison method 
the number of inequivalent MSSM-like orbifold models increases only by 3\%.

\subsection{Search in the ``OrbifoldLandscape''}
\label{sec:searchstrategy}

Using the improved version of the orbifolder we performed a scan in the landscape of all $\Z{N}$ 
and certain $\Z{N}\times\Z{M}$ heterotic orbifold geometries for MSSM-like models, where our 
basic requirements for a model to be MSSM-like are:
\begin{itemize}
\item SM gauge group $\SU{3}\times\SU{2}\times\U{1}_\text{Y} \subset \E{8}$ times a hidden sector
\item Hypercharge $\U{1}_\text{Y}$ is non-anomalous and in $\SU{5}$ GUT normalisation
\item (net) number of three generations of quarks and leptons
\item at least one Higgs pair
\item all exotics must be vector-like with respect to the SM gauge group
\end{itemize}
We identified approximately 12000 MSSM-like orbifold models that suit the above criteria. Given the 
large number of promising models we call them the ``OrbifoldLandscape''. A summary of the results can 
be found in the appendix in Tabs.~\ref{tab:MSSMsummary} and \ref{tab:MSSMsummary2}. Furthermore, the 
orbifolder input files needed to load these models into the program can be found at~\cite{WebTables:2014nv}. 
The scan did not reveal any MSSM-like models from orbifold geometries with $\Z{3}$, $\Z{7}$ and 
$\Z{2}\times\Z{6}$-II point group. This is most likely related to the condition of $\SU{5}$ GUT 
normalisation for hypercharge. 

Note that this search for MSSM-like orbifold models is by far not complete. For example, we only used 
the standard $\Z{N}\times\Z{M}$ orbifold geometries (i.e.\ those with label (1-1) following the 
nomenclature of Ref.~\cite{Fischer:2012qj}). In addition, our search was performed in a huge, but 
still finite parameter set of shifts and Wilson lines. Finally, the routine to identify inequivalent orbifold 
models can surely be improved further. Hence, presumably only a small fraction of the full heterotic 
orbifold Landscape has been analysed here.

\subsection{Comparison to the Literature}

Let us compare our findings to the literature. The $\Z{6}$-II (1-1) orbifold has been studied intensively 
in the past, see e.g.~\cite{Kobayashi:2004ya, Buchmuller:2005jr,Buchmuller:2006ik,Ratz:2007my}. Also the 
MiniLandscape~\cite{Lebedev:2006kn, Lebedev:2008un} was performed using this orbifold geometry, see 
section~\ref{sec:reviewminilandscape}. In the 
first paper~\cite{Lebedev:2006kn} local $\SO{10}$ and $\E{6}$ GUTs were used as a search strategy and 
thus one was restricted to four out of 61 possible shifts, resulting in 223 MSSM-like models. In the 
second paper~\cite{Lebedev:2008un} this restriction was lifted, resulting in almost 300 MSSM-like models. 
They are all included in our set of 348 MSSM-like models from $\Z{6}$-II (1-1), see Tab.~\ref{tab:MSSMsummary} 
in the appendix. 

Similar to $\Z{6}$-II, the $\Z{2}\times\Z{4}$ orbifold geometry has been conjectured to be very 
promising for MSSM model--building~\cite{Pena:2012ki}. Here, we can confirm this conjecture: we found 
3632 MSSM-like models from $\Z{2}\times\Z{4}$ (1-1) --- the largest set of models in our scan. Also from 
a geometrical point of view, the $\Z{2}\times\Z{4}$ orbifold is very rich: there are in total 41 different 
orbifold geometries with $\Z{2}\times\Z{4}$ point group, i.e.\ based on different six-tori and 
roto--translations~\cite{Fischer:2012qj}. We considered only the standard choice here, labelled (1-1). 
Hence, one can expect a huge landscape of MSSM-like models to be discovered from $\Z{2}\times\Z{4}$.
 
Recently, Groot Nibbelink and Loukas performed a model scan in all $\Z{8}$-I and $\Z{8}$-II 
geometries~\cite{Nibbelink:2013lua}. They also used a local GUT search strategy (based on $\SU{5}$ 
and $\SO{10}$ local GUTs) and hence started with 120 and 108 inequivalent shifts for $\Z{8}$-I and 
$\Z{8}$-II, respectively. Their scan resulted in 753 MSSM-like models. Without imposing the local GUT 
strategy our search revealed in total 1713 MSSM-like models from $\Z{8}$, see Tab.~\ref{tab:MSSMsummary}. 

Further orbifold MSSM-like models have been constructed using the $\Z{12}$-I orbifold 
geometry~\cite{Kim:2006hw,Kim:2007mt}. This orbifold seems also to be very promising as we identified 750 
MSSM-like models in this case, see Tab.~\ref{tab:MSSMsummary}. Finally, we confirm the analysis of 
Ref.~\cite{RamosSanchez:2008tn} for the orbifold geometries $\Z{6}$-I and $\Z{N}$ with $N=3,4,7$ and 
standard lattice (1-1).

In the next section we will apply the strategies described by the ``Five golden rules of superstring 
phenomenology'' to our OrbifoldLandscape and search for common properties of our 12000 MSSM-like 
orbifold models. Thereby, we will see how many MSSM-like models would have been found following the 
``Five golden rules'' strictly and how many would have been lost. Hence, we will estimate the prosperity 
of the ``Five golden rules''.

%%%%%%%%%%%%%%%%%%%%%%%%%%%%%%%%%%%%%%%%%%%%%%%%%%%%%%%%%%%%%%%%%%%%%%%%%%%%%%%%%%%%%%
\section{Five golden rules in the OrbifoldLandscape}
%%%%%%%%%%%%%%%%%%%%%%%%%%%%%%%%%%%%%%%%%%%%%%%%%%%%%%%%%%%%%%%%%%%%%%%%%%%%%%%%%%%%%%
\label{sec:rulesorbifoldlandscape}

In the following we focus on the golden rules I - IV. A detailed analysis of rule V is very model-dependent 
and will thus not be discussed here.

%%%%%%%%%%%%%%%%%%%%%%%%%%%%%%%%%%%%%%%%%%%%%%%%%%%%%%%%%%%%%%%%%%%%%%%%%%%%%%%%%%%%%%
\subsection{Rule I: Spinor of $\boldsymbol{\SO{10}}$ for SM matter}

\begin{figure}[t!]
\resizebox{0.99\textwidth}{!}{%
  \includegraphics{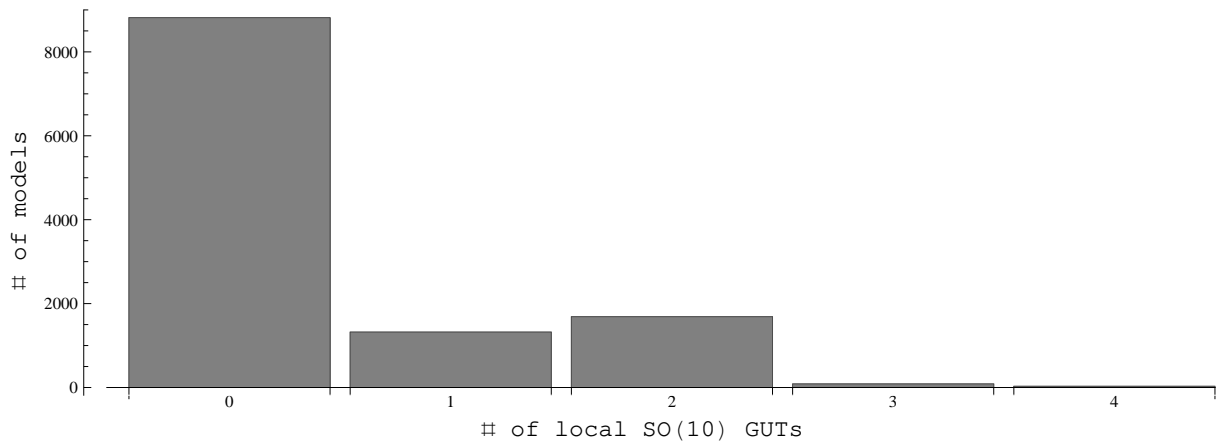}
}
\caption{Number of MSSM-like orbifold models vs. number of local $\SO{10}$ GUTs with $\rep{16}$-plets for matter.}
\label{fig:SO10Spinors}
\end{figure}

As discussed in Sec.~\ref{sec:firstgoldenrules} at least some generations of quarks and leptons might originate 
from spinors of $\SO{10}$ sitting at points in extra dimensions with local $\SO{10}$ GUT.\footnote{See 
also~\cite{Forste:2004ie, Buchmuller:2004hv} and for an overview on local GUTs Ref.~\cite{Ratz:2007my}.} 
Hence, we perform a statistic on the number of such localisations in our 12000 MSSM-like orbifold models. The 
results are summarised in Tab. \ref{tab:MSSMsummary2} and displayed in Fig.~\ref{fig:SO10Spinors}. 

It turns out that 25\% of our models have at least one local $\SO{10}$ GUT. Furthermore, we find that 
some orbifolds seem to forbid local $\SO{10}$ GUTs with $\rep{16}$-plets (for example $\Z{6}$-I~\cite{Ratz:2007my}). 
On the other hand, the MSSM-like models from $\Z{6}$-II and $\Z{8}$-I (1-1) and (2-1) prefer zero or two 
localised $\rep{16}$-plets of $\SO{10}$. Three local $\rep{16}$-plets are very uncommon, they mostly appear 
in $\Z{2}\times\Z{4}$.
 
Note that the number of local GUTs can be greater than three even though the model has a (net) 
number of three generations of quarks and leptons. Obviously, an anti-generation of quarks 
and leptons is needed in such a case. The maximal number we found in our scan is four local 
$\SO{10}$ GUTs with $\rep{16}$-plets for matter in the cases of $\Z{2}\times\Z{2}$ and 
$\Z{2}\times\Z{4}$ orbifold geometries.

\subsubsection{Other local GUTs}

\begin{figure}[t!]
\resizebox{0.99\textwidth}{!}{%
  \includegraphics{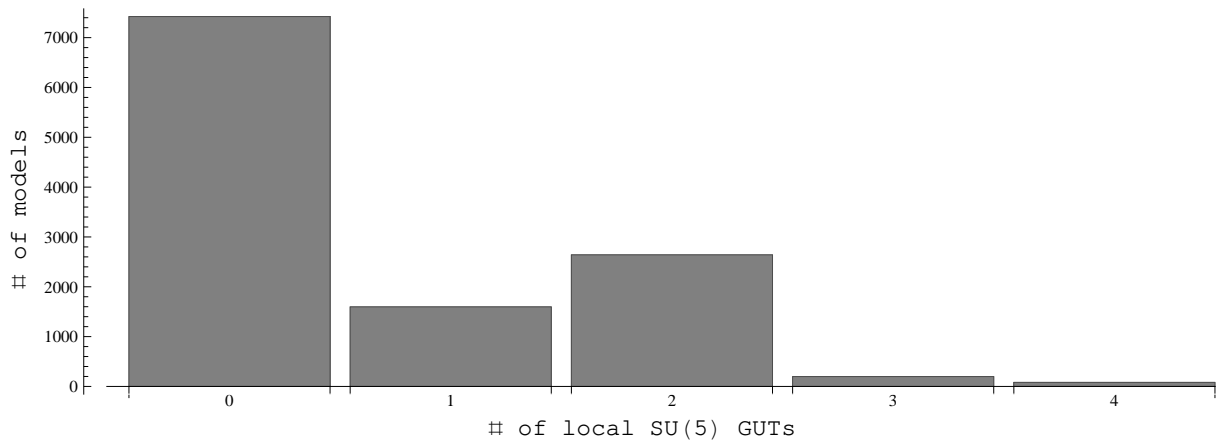}
}
\caption{Number of MSSM-like orbifold models vs. number of local $\SU{5}$ GUTs with $\rep{10}$-plets for matter.}
\label{fig:LocalSU5GUTs}
\end{figure}

In addition, we analyse our 12000 models for local $\SU{5}$ GUTs with local matter in $\rep{10}$-plets. The results 
are summarised in \Tabref{tab:MSSMsummary2} and displayed in Fig.~\ref{fig:LocalSU5GUTs}. We find this case to be 
very common: almost 40\% of our MSSM-like models have at least one local $\rep{10}$-plet of a local $\SU{5}$ GUT.

Next, we also look for local $\E{6}$ GUTs with $\rep{27}$-plets. We find only a few cases, most of them appear in 
$\Z{N}\times\Z{M}$ orbifold geometries, see \Tabref{tab:MSSMsummary2}. 

\begin{figure}[t]
\resizebox{0.99\textwidth}{!}{%
  \includegraphics{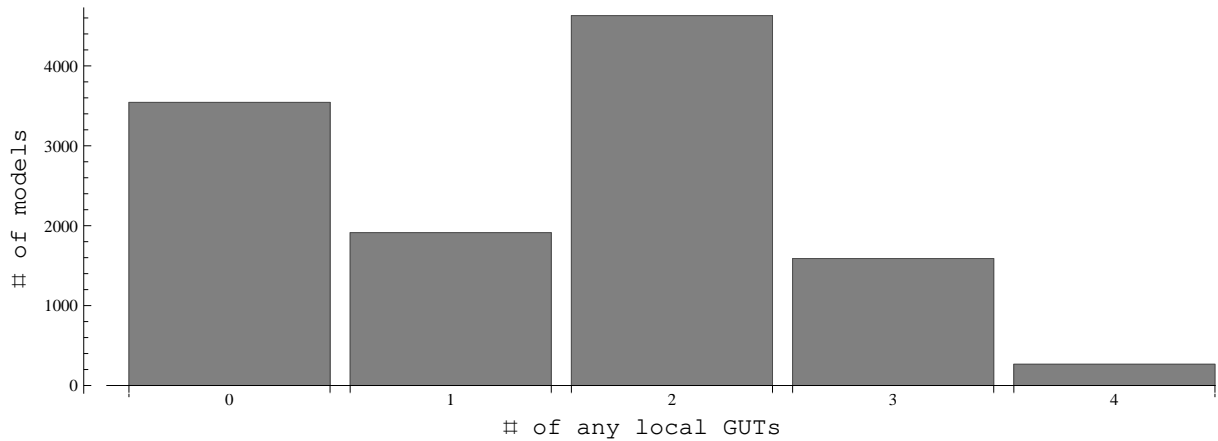}
}
\caption{Number of MSSM-like orbifold models vs. number of local GUTs with local GUT multiplets for SM matter.}
\label{fig:LocalAnyGUTs}
\end{figure}

Finally, we scan our models for localised SM generations (i.e.\ localised left--handed quark--doublets) transforming in a 
complete multiplet of any local GUT group that unifies the SM gauge group. Again, our results are listed in 
\Tabref{tab:MSSMsummary2} and visualised in Fig.~\ref{fig:LocalAnyGUTs}. We find most of our models, i.e.\ 70\%, have 
at least one local GUT with a localised SM generation. 

In summary, the first golden rule, which demands for local GUTs in extra dimensions in order to obtain complete GUT multiplets for matter, 
is very successful: most of our 12000 MSSM-like models share this property automatically, it was not imposed by hand in our search.

%%%%%%%%%%%%%%%%%%%%%%%%%%%%%%%%%%%%%%%%%%%%%%%%%%%%%%%%%%%%%%%%%%%%%%%%%%%%%%%%%%%%%%
\subsection{Rule II: Incomplete GUT Multiplets for the Higgs Pair}
\label{sec:orbifoldlandscaperule2}

\begin{figure}
\resizebox{0.99\textwidth}{!}{%
  \includegraphics{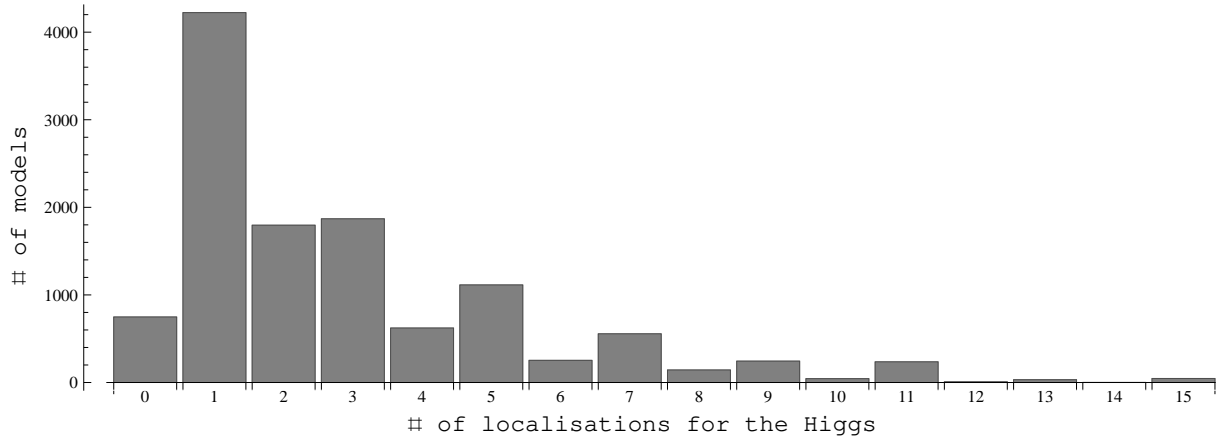}
}
\caption{Number of orbifold models vs. number of localisations with broken local GUT such that only a Higgs doublet but not 
the triplet survives. The 10D bulk is the most common localisation of this kind.}
\label{fig:Higgs}
\end{figure}

Since the Higgs doublets reside in incomplete GUT multiplets, they might be localised at some region of the orbifold where the 
higher-dimensional GUT is broken to the 4D Standard Model gauge group. This scenario yields a natural solution to the doublet--triplet 
splitting problem. The untwisted sector (i.e.\ bulk) would be a prime candidate for such a localisation, but there can be further 
possibilities. The numbers of such GUT breaking localisations are summarised in \Tabref{tab:MSSMsummary} and displayed in 
Fig.~\ref{fig:Higgs}.

We see that GUT breaking localisations are very common among our MSSM-like models. Only a very few models do not contain any GUT breaking 
localisations that yield incomplete GUT multiplets for at least one Higgs. On the other hand, there are 4223 cases with one GUT breaking localisation 
--- in most cases (4097 out of 4223) this is the bulk. In addition, there are many models that have more than one possibility for naturally 
split Higgs multiplets, but in almost all cases the bulk is among them. 

Note that most of our MSSM-like models have additional exotic Higgs-like pairs, mostly two to six additional ones. In contrast to the MSSM Higgs 
pair they often originate from complete multiplets of some local GUT. On the other hand, we identified 1011 MSSM-like models with exactly one 
Higgs pair. Cases with exactly one Higgs pair, originating from the bulk might be especially interesting.

In summary, the second golden rule, which explains incomplete GUT multiplets for the Higgs using GUT breaking localisations in extra dimensions, 
is very successful --- as in the case of the first golden rule, most of our 12000 MSSM-like models follow this rule automatically.

%%%%%%%%%%%%%%%%%%%%%%%%%%%%%%%%%%%%%%%%%%%%%%%%%%%%%%%%%%%%%%%%%%%%%%%%%%%%%%%%%%%%%%
\subsection{Rule III: Repetition of families}
\label{sec:orbifoldlandscaperule3}

The Standard Model contains three generations of quarks and leptons with a peculiar pattern of masses and mixings. This 
might be related to a (discrete) flavour symmetry.\footnote{A gauged flavour symmetry like $\SU{2}$ or $\SU{3}$ is also possible. 
Some of the models in our OrbifoldLandscape realise this possibility, but we do not analyse these cases in detail here.}

From the orbifold perspective discrete flavour symmetries naturally arise from the symmetries of the orbifold 
geometry~\cite{Kobayashi:2006wq,Nilles:2012cy}. However, certain background fields (i.e.\ orbifold Wilson lines~\cite{Ibanez:1986tp}) 
can break these symmetries. The maximal number of orbifold Wilson lines is six corresponding to the six directions of the compactified 
space. The orbifold--rotation, however, in general identifies some of those directions. Hence, the corresponding Wilson lines have to be 
equal. For example, the $\Z{3}$ orbifold allows for maximally three independent Wilson lines.

In general, one can say that the more Wilson lines vanish the larger is the discrete flavour symmetry. On the other hand, 
non-vanishing Wilson lines are generically needed in order to obtain the Standard Model gauge group and to reduce the number of 
generations to three. Hence, it is interesting to perform a statistic on the number of vanishing Wilson lines for our 12000 MSSM-like 
orbifold models, see Tab.~\ref{tab:MSSMsummary} in the appendix and Fig.~\ref{fig:Generations}.

\begin{figure}
\resizebox{0.99\textwidth}{!}{%
  \includegraphics{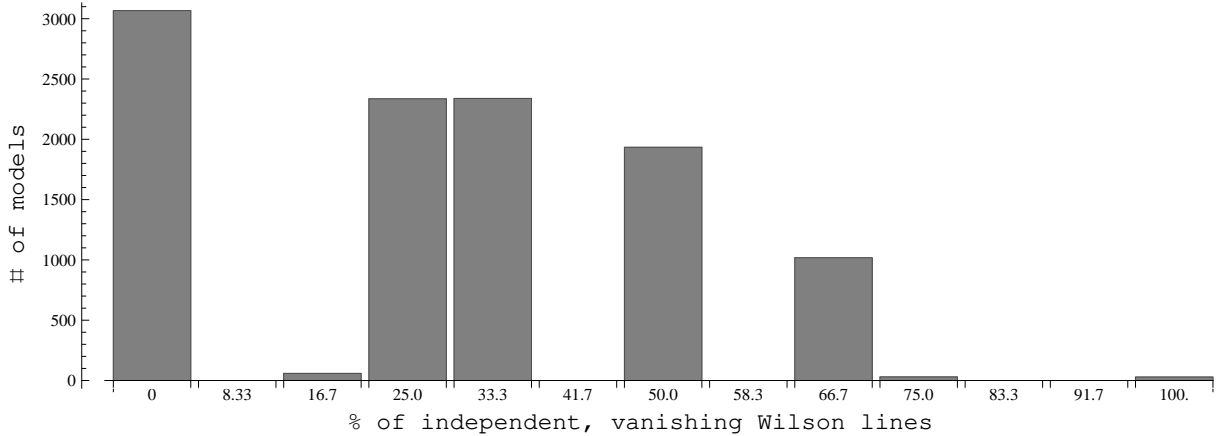}
}
\caption{Number of MSSM-like orbifold models vs. the percentage of independent Wilson lines that are vanishing (e.g.\ the 
$\Z{2}\times\Z{2}$ orbifold allows for six independent Wilson lines. If one is vanishing, the percentage is 16.7\%). Higher 
percentages generically correspond to larger flavour symmetries.}
\label{fig:Generations}
\end{figure}

There are orbifold geometries, like $\Z{4}$, $\Z{6}$-I and $\Z{12}$-I, apparently demanding for all possible orbifold Wilson lines 
to be non-trivial in order to yield the MSSM, see Tab.~\ref{tab:MSSMsummary}. These MSSM-like models are expected to have no 
discrete, non-Abelian flavour symmetries. On the other hand, there are several orbifold geometries 
that seem to require at least one vanishing Wilson line in order to reproduce the MSSM with its three generations, for example $\Z{6}$-II, 
$\Z{2}\times\Z{2}$, $\Z{2}\times\Z{4}$, $\Z{3}\times\Z{3}$ and $\Z{4}\times\Z{4}$. In general, the case of vanishing Wilson lines 
is very common: we see that in 75\% of our MSSM-like orbifold models at least one allowed orbifold Wilson line is zero. In these cases 
non-Abelian flavour symmetries are expected. For example, most of the MSSM-like models from $\Z{6}$-II (1-1) have a $D_4$ flavour 
symmetry with the first two generations transforming as a doublet and the third one as a 
singlet~\cite{Kobayashi:2004ya,Kobayashi:2006wq,Nilles:2012cy}.

In summary, the third golden rule, which explains the origin of three generations of quarks and leptons by geometrical properties of the 
compactification space, is generically satisfied for our 12000 MSSM-like orbifold models.

%%%%%%%%%%%%%%%%%%%%%%%%%%%%%%%%%%%%%%%%%%%%%%%%%%%%%%%%%%%%%%%%%%%%%%%%%%%%%%%%%%%%%%
\subsection{Rule IV: $\boldsymbol{\mathcal{N} = 1}$ supersymmetry}
\label{sec:orbifoldlandscaperule4}

By construction, i.e.\ by choosing the appropriate orbifold geometries, our 12000 MSSM-like orbifold models preserve 
$\mathcal{N}=1$ supersymmetry in four dimensions. This is expected to be broken by non-perturbative effects, i.e.\ 
by hidden sector gaugino condensation~\cite{Nilles:1982ik,Ferrara:1982qs,Derendinger:1985kk,Dine:1985rz}. 
Here, we follow the discussion of~\cite{Lebedev:2006tr} where low energy 
supersymmetry breaking in the MiniLandscape of $\Z{6}$-II orbifolds was analysed. See also~\cite{Dienes:2006ut,Dienes:2007ms} 
for a related discussion. 

In detail, our MSSM-like models typically possess a non-Abelian hidden sector 
gauge group with little or no charged matter representations. The corresponding gauge coupling depends via the one-loop 
$\beta$-functions on the energy scale. If the coupling becomes strong at some (intermediate) energy scale $\Lambda$ the 
respective gauginos condensate and supersymmetry is broken spontaneously by a non-vanishing dilaton $F$-term. 
Assuming that SUSY breaking is communicated 
to the observable sector via gravity the scale of soft SUSY breaking is given by the gravitino mass, i.e.\
\begin{equation}
m_{3/2} \,\sim\, \frac{\Lambda^3}{M_\text{Plank}^2}\;,
\end{equation}
where $M_\text{Plank}$ denotes the Planck mass and the scale of gaugino condensation $\Lambda$ is given by
\begin{equation}
\Lambda \,\sim\, M_\text{GUT}\, \text{exp}\left(-\frac{1}{2 \beta} \frac{1}{g^2(M_\text{GUT})}\right)\;.
\end{equation}

\begin{figure}
\resizebox{0.99\textwidth}{!}{%
  \includegraphics{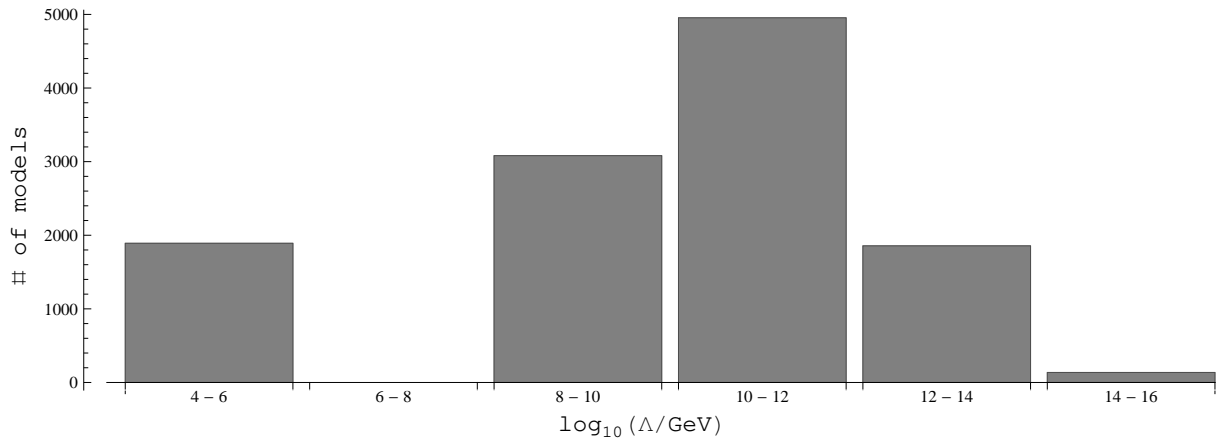}
}
\caption{Number of MSSM-like orbifold models vs. scale of gaugino condensation for the largest hidden sector gauge group.}
\label{fig:GauginoCondenstaion}
\end{figure}

For every MSSM-like orbifold model we compute the $\beta$-function of the largest hidden sector gauge group under the assumption 
that any non-trivial hidden matter representation of this gauge group can be decoupled in a supersymmetric way. Furthermore, we 
assume dilaton stabilization at a realistic value $1/g^2(M_\text{GUT}) = \text{Re}S \approx 2$. Hence, we obtain the scale 
$\Lambda$ of gaugino condensation. Our results are displayed in Fig.~\ref{fig:GauginoCondenstaion}. For an intermediate scale 
$\Lambda \sim 10^{13} \text{GeV}$ one obtains a gravitino mass in the $\text{TeV}$ range, which is of phenomenological interest. 

The models in the OrbifoldLandscape seem to prefer low energy SUSY breaking. This result is strongly related to the 
heterotic orbifold construction: the $\E{8}\times\E{8}$ gauge group in 10D is broken by orbifold shift and Wilson lines, which are 
highly constrained by string theory (i.e.\ modular invariance). Therefore, both $\E{8}$ factors get broken and not only the 
observable one. It turns out that the unbroken gauge group from the hidden $\E{8}$ has roughly the correct size to yield 
gaugino condensation at an intermediate scale and hence low energy SUSY breaking.

Note that our analysis is just a rough estimate as various effects have been neglected, for example the decoupling of hidden 
matter, the identification of the gaugino condensation and (string) threshold corrections. These effects can in principle 
affect the scale of SUSY breaking even by 2-3 orders of magnitude.

%%%%%%%%%%%%%%%%%%%%%%%%%%%%%%%%%%%%%%%%%%%%%%%%%%%%%%%%%%%%%%%%%%%%%%%%%%%%%%%%%%%%%%
\section{The general Landscape}
%%%%%%%%%%%%%%%%%%%%%%%%%%%%%%%%%%%%%%%%%%%%%%%%%%%%%%%%%%%%%%%%%%%%%%%%%%%%%%%%%%%%%%
\label{sec:generallandscape}

With these considerations we have only scratched the surface of the parameter space of 
potentially realistic models. In addition, we have used ``five golden rules'' as a prejudice 
for model selection and it has to be seen whether this is really justified. 

For general model building in the framework of (perturbative) string theory we have the 
following theories at our disposal:
\begin{itemize} 
\item type I string with gauge group $\SO{32}$
\item heterotic $\SO{32}$
\item heterotic $\E{8}\times\E{8}$
\item type IIA and IIB orientifolds
\item intersecting branes with gauge group $\U{N}^M$
\end{itemize}
As we explained in detail, our rule I points towards exceptional groups and hence towards 
the $\E{8}\times\E{8}$ heterotic string. On the other hand, type II orientifolds typically 
provide gauge groups of type $\SO{M}$ or $\U{N}$ and products thereof. Although we have 
$\SO{2N}$ gauge groups in these schemes, matter fields do not come as spinors of $\SO{2N}$, 
but originate from adjoint representations. In the intersecting brane models based on 
$\U{N}^M$ gauge groups matter transforms in bifundamental representations of $\U{N}\times\U{L}$ 
(originating from the adjoint of $\U{N+L}$). While this works nicely for the standard 
model representations, it appears to be difficult to describe a grand unified picture with 
e.g.\ gauge group $\SU{5}$. Trying to obtain a GUT yields a gauge group at least as large as 
$\U{5}$ and one has problems with a perturbative top-quark Yukawa coupling. One possible way 
out is the construction of string models without the prejudice for GUTs, see 
e.g.~\cite{Gato-Rivera:2014afa}.

A comprehensive review on these intersecting brane model constructions can be found in 
the book of Ib{\'a}{\~n}ez and Uranga~\cite{Ibanez:2012zz} or other reviews~\cite{Blumenhagen:2006ci}. 
These models have a very appealing geometric interpretation, see e.g.\cite{Honecker:2012qr}: 
Fields are located on branes of various dimensions. Thus, physical properties of the models can be 
inferred from the localisation of the brane--fields in the extra dimensions and by the overlap of 
their wave functions, similar to the heterotic MiniLandscape. This nice geometrical set-up leads 
to attempts to construct so-called ``local models''. Here, one assumes that all particle physics 
properties of the model are specified by some local properties at some specific point or sub-space 
of the compactified dimensions and that the ``bulk'' properties can be decoupled. However, the 
embedding of the local model into an ultraviolet complete and consistent string model is an
assumption and its validity remains an open question.

Further schemes include ``non-perturbative'' string constructions:
\begin{itemize} 
\item M-theory in $D=11$
\item heterotic M-theory $\E{8}\times\E{8}$ 
\item F-theory
\end{itemize}
These non-perturbative constructions are conjectured theories that generalize string theories or known 
supergravity field theories in higher dimensions. The low energy limit of M-theory is 11-dimensional 
supergravity. Heterotic M-theory is based on a $D=11$ theory bounded by two $D=10$ branes with gauge 
group $\E{8}$ on each boundary and F-theory is a generalization of type IIB theory, where certain 
symmetries can be understood geometrically. This non-perturbative construction allows for 
singularities in extra dimensions that lead to non-trivial gauge groups according to the so-called A-D-E 
classification. Groups of the A-type ($\SU{N}$) and D-type ($\SO{2N}$) can also be obtained in the 
perturbative constructions with D-branes and orientifold branes, while exceptional gauge groups can only 
appear through the presence of E-type singularities. This allows for spinors of $\SO{10}$ and can 
produce a non-trivial top-quark Yukawa coupling within an $\SU{5}$ grand unified theory. In that sense, 
F-theory can be understood as an attempt to incorporate rule I within type IIB theory. Unfortunately, 
it is difficult to control the full non-perturbative theory and the search for realistic models is often 
based on local model building. Many questions are still open but there is enough room for optimism that 
promising models can be embedded in a consistent ultraviolet completion.

A general problem of string phenomenology is the difficulty to perform the explicit calculations needed to 
check the validity of the model. This is certainly true for the non-perturbative models, where we have 
(at best) some effective supergravity description. But also in the perturbative constructions we 
have to face this problem. We have to use simplified compactification schemes to be able to do the 
necessary calculations --- we need a certain level of ``Berechenbarkeit''. In our discussion we used 
the flat orbifold compactification that allows the use of conformal field theory methods. In principle, 
this enables us to do all the necessary calculations to check the models in detail. In the $\Z{6}$-II 
MiniLandscape this has been elaborated to a large extend. For the more general orbifold landscape, 
this still has to be done. Other constructions with full conformal field theory control are the free 
fermionic constructions~\cite{Faraggi:1991jr} and the ``tensoring'' of conformal field 
theory building blocks: so-called Gepner models~\cite{Dijkstra:2004cc}. They share ``Berechenbarkeit'' with the flat 
orbifold models, but the geometric structure of compactified space is less transparent. 

We have to hope that these simplified compactifications (or approximations) lead us to realistic models. 
In the generic situation one needs smooth manifolds, e.g. Calabi-Yau spaces, and some specific models have 
been constructed~\cite{Donagi:1999ez,Braun:2005ux}. However, these more generic compactifications require 
more sophisticated methods for computations that are only partially available, for example in order to 
determine Yukawa couplings. More recently a simplification based on the embedding of line bundles
has allowed the constructions of many models~\cite{Anderson:2011ns,Anderson:2012yf}. Still the calculational 
options are limited. It would be interesting to get a better geometric understanding of the compact manifold.
At the moment the ``determination'' of couplings is based on a supergravity approximation using $\U{1}$ 
symmetries. These symmetries are exact in this approximation at the ``stability wall'' but are expected to be 
broken to discrete symmetries in the full theory. This is in concord with rule V asking for the origin of 
discrete symmetries. Furthermore, this question has recently been analysed intensively within the various string 
constructions~\cite{BerasaluceGonzalez:2011wy,Ibanez:2012wg,BerasaluceGonzalez:2012vb, Anastasopoulos:2012zu,Honecker:2013hda}.

%%%%%%%%%%%%%%%%%%%%%%%%%%%%%%%%%%%%%%%%%%%%%%%%%%%%%%%%%%%%%%%%%%%%%%%%%%%%%%%%%%%%%%
\section{Summary}
%%%%%%%%%%%%%%%%%%%%%%%%%%%%%%%%%%%%%%%%%%%%%%%%%%%%%%%%%%%%%%%%%%%%%%%%%%%%%%%%%%%%%%
\label{sec:nvconclusions}

We have seen that there is still a long way to go in the search for realistic particle 
physics models from string theory. There are many possible roads but we are limited by 
our calculational techniques. Thus, in the near future we are still forced to make choices. 
Here, we have chosen to follow ``five golden rules'' outlined in section~\ref{sec:fivegoldenrules}, 
which are mainly motivated by the quest for a unified picture of particle physics interactions. 
This strategy seems to require an underlying structure provided by exceptional groups pointing
towards the $\E{8}\times\E{8}$ heterotic string and F-theory.

Even given these rules, there are stumbling blocks because of the complexity of the compact 
manifolds. We cannot resolve these problems in full generality: we have to use simplified 
compactification schemes or approximations. We have to hope that nature has chosen a theory 
that is somewhat close to these simplified schemes. Of course, any method to go beyond this 
simplified assumptions should be seriously considered. However, there is some hope that this 
assumption might be justified: The orbifold models studied in this work have enhanced (discrete) 
symmetries that could be the origin of symmetries of the standard model, especially with 
respect to the flavour structure and symmetries relevant for proton stability as well as the 
absence of other rare processes. Generically, these symmetries are slightly broken as we 
go away from the orbifold--point. This gives rise to some hierarchical structures, for 
example for the ratio of quark masses in the spirit of Froggatt and Nielsen~\cite{Froggatt:1978nt}.

The analysis of the MiniLandscape can be seen as an attempt to study these questions in 
detail. Based on the availability of conformal field theory techniques we can go pretty 
far in the analysis of explicit models. A detailed analysis of the ``OrbifoldLandscape'' 
has not been performed yet, but should be possible along the same lines. In 
section~\ref{sec:orbifoldlandscape} we started this enterprise of model building by 
constructing 12000 MSSM-like models. In a next step, the detailed properties of promising 
models have to be worked out. Especially the framework of the $\Z{2}\times\Z{4}$~\cite{Pena:2012ki} 
should provide new insight into the properties of realistic models and might teach us further 
key properties shared by successful models.

One key property that we have learned is the geography of fields in the extra dimensions. 
The localisation of matter fields and the gauge group profiles in extra dimensions are 
essential for the properties of the low energy model. This is the first message of the 
heterotic orbifold construction and shared by the ``braneworld'' constructions in type II 
string theory and F-theory. Further lessons are:
\begin{itemize}
\item The Higgs pair is a bulk field. This allows for a convincing solution of the 
$\mu$-problem using a (discrete) R-symmetry and yields doublet--triplet splitting.
\item A sizeable value of the top-quark Yukawa coupling requires a sufficient overlap 
with the Higgs fields in extra dimensions. Thus, the top-quark should extend to the bulk as well.
\item The matter fields of the first and second generation should be localised in a region 
of the extra-dimensional space where they are subject to an enhanced gauge symmetry, like $\SO{10}$. 
This local GUT forces them to appear as complete representations, e.g.\ as spinors of $\SO{10}$. 
Furthermore, the geometrical structure can manifest itself in a discrete flavour symmetry.
\item The quest for low energy supersymmetry is the guiding principle in string model building. 
Still, it has to be seen whether this is realised in nature. At the moment no sign of 
supersymmetry has been found at the LHC, although the value of the Higgs mass is consistent 
with SUSY. The analysis of the models of the MiniLandscape and the location of the
fields suggests a certain structure where even some remnants of
extended supersymmetry (for fields in the bulk) seem to be at work.
This picture of ``heterotic supersymmetry''~\cite{Krippendorf:2012ir,Badziak:2012yg} 
can hopefully be tested experimentally in the not too far future.
\end{itemize}

\subsection*{Acknowledgments}

This work was partially supported by the SFB--Transregio TR33 ``The Dark
Universe'' (Deutsche Forschungsgemeinschaft)
and 
the DFG cluster of excellence ``Origin and Structure of the Universe'' (www.universe-cluster.de).

%%%%%%%%%%%%%%%%%%%%%%%%%%%%%%%%%%%%%%%%%%%%%%%%%%%%%%%%%%%%%%%%%%%%%%%%%%%%%%%%%%%%%%
% Appendix
%%%%%%%%%%%%%%%%%%%%%%%%%%%%%%%%%%%%%%%%%%%%%%%%%%%%%%%%%%%%%%%%%%%%%%%%%%%%%%%%%%%%%%

\appendix

\section{Summary of the OrbifoldLandscape}

\begin{landscape}
\begin{table}[ht]
\begin{tabular}{@{}cc|r||r|rrrrr|rrrrrrr|r@{}}
\hline
                      &      &                   &\multicolumn{1}{|c|}{max. \# } &\multicolumn{5}{|c|}{\# models with}      &\multicolumn{7}{|c}{\# models with}           & \multicolumn{1}{|c}{\!\!\# MSSM\!\!}\\
\multicolumn{2}{c|}{orbifold}&\!\!\!\# MSSM\!\!\!&\multicolumn{1}{|c|}{of indep.}&0 & 1 & 2 & 3 & $\geq4$                   &0 & 1 & 2 & 3 & 4 & 5 & $\geq 6$              & \multicolumn{1}{|c}{\!\!without\!\!}\\
                      &      &                   &\multicolumn{1}{|c|}{WLs}      &\multicolumn{5}{|c|}{indep. vanishing WLs}&\multicolumn{7}{|c}{locations for split Higgs}& \multicolumn{1}{|c}{\!\!$\U{1}_\text{anom}$\!\!}\\
\hline
$\Z{3}$               & (1,1) &    0 & 3 &   0 &    0 &    0 &  0 & 0 &   0 &   0 &   0 &   0 &   0 &   0 &   0 &   0\\
\hline
$\Z{4}$               & (1,1) &    0 & 4 &   0 &    0 &    0 &  0 & 0 &   0 &   0 &   0 &   0 &   0 &   0 &   0 &   0\\
                      & (2,1) &  128 & 3 & 128 &    0 &    0 &  0 & 0 &   6 & 107 &  12 &   3 &   0 &   0 &   0 &   0\\
                      & (3,1) &   25 & 2 &  25 &    0 &    0 &  0 & 0 &   0 &  25 &   0 &   0 &   0 &   0 &   0 &   0\\
\hline
$\Z{6}$-I             & (1,1) &   31 & 1 &  31 &    0 &    0 &  0 & 0 &   0 &  31 &   0 &   0 &   0 &   0 &   0 &   0\\
                      & (2,1) &   31 & 1 &  31 &    0 &    0 &  0 & 0 &   0 &  31 &   0 &   0 &   0 &   0 &   0 &   0\\
\hline
$\Z{6}$-II            & (1,1) &  348 & 3 &  13 &  335 &    0 &  0 & 0 &  20 & 167 & 111 &  34 &   8 &   2 &   6 &   1\\
                      & (2,1) &  338 & 3 &  10 &  328 &    0 &  0 & 0 &  19 & 162 & 107 &  33 &   9 &   2 &   6 &   2\\
                      & (3,1) &  350 & 3 &  18 &  332 &    0 &  0 & 0 &  17 & 172 & 112 &  41 &   7 &   1 &   0 &   2\\
                      & (4,1) &  334 & 2 &  39 &  295 &    0 &  0 & 0 &  17 & 161 & 113 &  32 &  11 &   0 &   0 &   3\\
\hline
$\Z{7}$               & (1,1) &    0 & 1 &   0 &    0 &    0 &  0 & 0 &   0 &   0 &   0 &   0 &   0 &   0 &   0 &   0\\
\hline
$\Z{8}$-I             & (1,1) &  263 & 2 & 221 &   42 &    0 &  0 & 0 &   0 & 128 &  85 &  50 &   0 &   0 &   0 &   7\\
                      & (2,1) &  164 & 2 & 123 &   41 &    0 &  0 & 0 &   0 &  76 &  53 &  35 &   0 &   0 &   0 &   5\\
                      & (3,1) &  387 & 1 & 387 &    0 &    0 &  0 & 0 &  27 & 150 & 175 &  32 &   3 &   0 &   0 &  27\\
\hline
$\Z{8}$-II            & (1,1) &  638 & 3 & 212 &  404 &   22 &  0 & 0 &  12 & 257 & 165 & 123 &  16 &  50 &  15 &   7\\
                      & (2,1) &  260 & 2 &  92 &  168 &    0 &  0 & 0 &  15 & 108 &  84 &  34 &   2 &  12 &   5 &   3\\
\hline
$\Z{12}$-I            & (1,1) &  365 & 1 & 365 &    0 &    0 &  0 & 0 &   5 & 259 &  55 &  42 &   4 &   0 &   0 &   8\\
                      & (2,1) &  385 & 1 & 385 &    0 &    0 &  0 & 0 &   7 & 271 &  63 &  44 &   0 &   0 &   0 &   9\\
\hline
$\Z{12}$-II           & (1,1) &  211 & 2 & 135 &   76 &    0 &  0 & 0 &   9 &  40 & 107 &  31 &  12 &   4 &   8 &   3\\
\hline
$\Z{2}\times\Z{2}$    & (1,1) &  101 & 6 &   0 &   59 &   42 &  0 & 0 &  79 &   0 &  10 &   3 &   8 &   0 &   1 &   0\\
\hline
$\Z{2}\times\Z{4}$    & (1,1) & 3632 & 4 &  67 & 2336 & 1199 & 30 & 0 & 393 &1194 & 160 & 690 &  83 & 449 & 663 &  10\\
\hline
$\Z{2}\times\Z{6}$-I  & (1,1) &  445 & 2 & 332 &  113 &    0 &  0 & 0 &  54 & 118 & 105 &  79 &  27 &  13 &  49 &   5\\
\hline
$\Z{2}\times\Z{6}$-II & (1,1) &    0 & 0 &   0 &    0 &    0 &  0 & 0 &   0 &   0 &   0 &   0 &   0 &   0 &   0 &   0\\
\hline
$\Z{3}\times\Z{3}$    & (1,1) &  445 & 3 &   1 &  369 &   75 &  0 & 0 &  27 & 212 &   1 &  15 & 102 &   0 &  88 &   9\\
\hline
$\Z{3}\times\Z{6}$    & (1,1) &  465 & 1 & 441 &   24 &    0 &  0 & 0 &   4 &  39 &  64 &  82 &  88 & 110 &  78 &   0\\
\hline
$\Z{4}\times\Z{4}$    & (1,1) & 1466 & 3 &  11 &  529 &  921 &  5 & 0 &  28 & 441 &  49 & 195 &  81 & 323 & 349 &   1\\
\hline
$\Z{6}\times\Z{6}$    & (1,1) & 1128 & 0 &1128 &    0 &    0 &  0 & 0 &   9 &  74 & 165 & 271 & 161 & 148 & 300 &   0\\
\hline
\hline
total                 &       & 11940&   &     &      &      &    &   & 748 &4223 &1796 &1869 & 622 &1114 &1568 &102 \\
\hline
\end{tabular}
\caption{\scriptsize Statistics on MSSM-like models (using the search criteria listed in Sec.~\ref{sec:searchstrategy}) obtained from a random scan in all $\Z{N}$ and certain $\Z{N}\times\Z{M}$ heterotic orbifold geometries. The first column labels the geometry following the nomenclature from~\cite{Fischer:2012qj}. The second column gives the number of inequivalent MSSM-like models found in our scan. Next, we 
give the maximal number of independent Wilson lines (WLs) possible for the respective orbifold geometry and in the fourth column we count the number of MSSM-like models with a certain number (i.e.\ 0,1,2,3,4) of vanishing Wilson lines, see Sec. \ref{sec:orbifoldlandscaperule3}. In the fifth column we count the number of locations with broken local GUT such that Higgs-doublets without triplets appear, see Sec. \ref{sec:orbifoldlandscaperule2}. Finally, in the last column we give the number of models without $\U{1}_\text{anom}$, i.e.\ without FI term.}
\label{tab:MSSMsummary}
\end{table}
\end{landscape}

\begin{landscape}
\begin{table}[ht]
\begin{tabular}{@{}cc||rrrrr|rrr|rrrrr||rrrrr@{}}
\hline
                       &      &\multicolumn{5}{|c|}{\# models with}      &\multicolumn{3}{|c|}{\# models with}    &\multicolumn{5}{|c||}{\# models with}     &\multicolumn{5}{|c}{\# models with}\\
\multicolumn{2}{c||}{orbifold}&0 & 1 & 2 & 3 & 4                         &0 & 1 & 2                               &0 & 1 & 2 & 3 & 4                         &0 & 1 & 2 & 3 & 4\\
                       &      &\multicolumn{5}{|c|}{local $\SO{10}$ GUTs}&\multicolumn{3}{|c|}{local $\E{6}$ GUTs}&\multicolumn{5}{|c||}{local $\SU{5}$ GUTs}&\multicolumn{5}{|c}{local GUTs}\\
\hline
$\Z{3}$               & (1,1) &    0 &   0 &   0 &   0 &  0 &   0 &   0 &   0 &    0 &   0 &   0 &   0 &  0 &   0 &   0 &   0 &   0 &  0\\
\hline
$\Z{4}$               & (1,1) &    0 &   0 &   0 &   0 &  0 &   0 &   0 &   0 &    0 &   0 &   0 &   0 &  0 &   0 &   0 &   0 &   0 &  0\\
                      & (2,1) &   78 &  50 &   0 &   0 &  0 &  50 &  78 &   0 &  128 &   0 &   0 &   0 &  0 &   0 & 128 &   0 &   0 &  0\\
                      & (3,1) &    5 &  20 &   0 &   0 &  0 &  20 &   5 &   0 &   25 &   0 &   0 &   0 &  0 &   0 &  25 &   0 &   0 &  0\\
\hline
$\Z{6}$-I             & (1,1) &   31 &   0 &   0 &   0 &  0 &  31 &   0 &   0 &   31 &   0 &   0 &   0 &  0 &  31 &   0 &   0 &   0 &  0\\
                      & (2,1) &   31 &   0 &   0 &   0 &  0 &  31 &   0 &   0 &   31 &   0 &   0 &   0 &  0 &  31 &   0 &   0 &   0 &  0\\
\hline
$\Z{6}$-II            & (1,1) &  155 &   2 & 187 &   4 &  0 & 332 &   6 &  10 &  203 &  12 & 133 &   0 &  0 &   2 &   3 & 293 &   4 & 46\\
                      & (2,1) &  148 &   1 & 186 &   3 &  0 & 323 &   5 &  10 &  204 &   6 & 128 &   0 &  0 &   2 &   5 & 324 &   4 &  3\\
                      & (3,1) &  164 &   1 & 185 &   0 &  0 & 328 &  11 &  11 &  202 &  12 & 136 &   0 &  0 &   2 &  11 & 293 &   9 & 35\\
                      & (4,1) &  158 &   3 & 173 &   0 &  0 & 299 &  23 &  12 &  195 &  18 & 121 &   0 &  0 &   0 &  14 & 315 &   5 &  0\\
\hline
$\Z{7}$               & (1,1) &    0 &   0 &   0 &   0 &  0 &   0 &   0 &   0 &    0 &   0 &   0 &   0 &  0 &   0 &   0 &   0 &   0 &  0\\
\hline
$\Z{8}$-I             & (1,1) &  143 &   0 & 120 &   0 &  0 & 263 &   0 &   0 &  226 &  37 &   0 &   0 &  0 & 106 &  31 & 120 &   6 &  0\\
                      & (2,1) &   92 &   0 &  72 &   0 &  0 & 164 &   0 &   0 &  147 &  17 &   0 &   0 &  0 &  75 &  15 &  74 &   0 &  0\\
                      & (3,1) &  164 & 140 &  83 &   0 &  0 & 346 &  32 &   9 &  336 &  29 &  22 &   0 &  0 & 105 & 117 & 133 &  32 &  0\\
\hline
$\Z{8}$-II            & (1,1) &  428 &  77 & 133 &   0 &  0 & 638 &   0 &   0 &  276 & 155 & 207 &   0 &  0 &  79 & 194 & 355 &  10 &  0\\
                      & (2,1) &  180 &  29 &  51 &   0 &  0 & 260 &   0 &   0 &   89 &  52 & 114 &   5 &  0 &  28 &  29 & 185 &  18 &  0\\
\hline
$\Z{12}$-I            & (1,1) &  365 &   0 &   0 &   0 &  0 & 259 &   0 & 106 &  365 &   0 &   0 &   0 &  0 & 250 &   0 & 115 &   0 &  0\\
                      & (2,1) &  385 &   0 &   0 &   0 &  0 & 269 &   0 & 116 &  385 &   0 &   0 &   0 &  0 & 269 &   0 & 116 &   0 &  0\\
\hline
$\Z{12}$-II           & (1,1) &  110 &  69 &  32 &   0 &  0 & 177 &  31 &   3 &   86 &  78 &  47 &   0 &  0 &   0 &  80 & 131 &   0 &  0\\
\hline
$\Z{2}\times\Z{2}$    & (1,1) &   72 &   6 &  12 &   1 & 10 &  66 &  33 &   2 &   75 &   0 &  11 &   0 & 15 &   3 &  18 &   8 &  30 & 42\\
\hline
$\Z{2}\times\Z{4}$    & (1,1) & 2948 & 300 & 297 &  68 & 19 &3181 & 358 &  93 &\!\!\!\!\!2831 &  71 & 707 &   7 & 16 &\!\!\!\!\!1918 &  70 & 670 & 911 & 63\\
\hline
$\Z{2}\times\Z{6}$-I  & (1,1) &  312 & 124 &   9 &   0 &  0 & 252 &  63 & 130 &  245 & 126 &  71 &   3 &  0 &  40 &  66 & 193 & 119 & 27\\
\hline
$\Z{2}\times\Z{6}$-II & (1,1) &    0 &   0 &   0 &   0 &  0 &   0 &   0 &   0 &    0 &   0 &   0 &   0 &  0 &   0 &   0 &   0 &   0 &  0\\
\hline
$\Z{3}\times\Z{3}$    & (1,1) &  444 &   1 &   0 &   0 &  0 & 445 &   0 &   0 &  289 &   3 &   2 & 151 &  0 & 246 &   2 &   3 & 194 &  0\\
\hline
$\Z{3}\times\Z{6}$    & (1,1) &  396 &  33 &  36 &   0 &  0 & 463 &   2 &   0 &   77 & 294 &  42 &  12 & 40 &   3 & 291 & 116 &  15 & 40\\
\hline
$\Z{4}\times\Z{4}$    & (1,1) & 1246 & 116 &  94 &  10 &  0 &1293 & 173 &   0 &  703 &  31 & 709 &  13 & 10 & 353 & 205 & 674 & 224 & 10\\
\hline
$\Z{6}\times\Z{6}$    & (1,1) &  761 & 349 &  18 &   0 &  0 &1122 &   6 &   0 &  274 & 656 & 191 &   7 &  0 &   0 & 609 & 511 &   8 &  0\\
\hline
\hline
total                 &       & 8816 &\!\!\!\!\!1321 &\!\!\!\!\!1688 &  86 & 29 &\!\!\!\!\!10612&826 & 502 &\!\!\!\!\!7423 &1597 &2641 & 198 & 81 &\!\!\!\!\!3543 &1913 &4629 &1589 &266\\
\hline
\end{tabular}
\caption{\scriptsize Statistics on MSSM-like models (using the search criteria listed in Sec.~\ref{sec:searchstrategy}) obtained from a random scan in all $\Z{N}$ and certain $\Z{N}\times\Z{M}$ heterotic orbifold geometries. The first column labels the geometry following the nomenclature from~\cite{Fischer:2012qj}. The next four columns display the number of MSSM-like models with 0,1,2,3 and (up to) 4 local GUTs of specified gauge group with corresponding local matter: local $\SO{10}$ GUTs with local $\rep{16}$-plets, local $\E{6}$ GUTs with local $\rep{27}$-plets, local $\SU{5}$ GUTs with local $\rep{10}$-plets and, finally, any local GUTs that unify $\SU{3}\times\SU{2}\times\U{1}_\text{Y}$ in a single gauge group with corresponding local matter representations containing left--handed quark doublets.}
\label{tab:MSSMsummary2}
\end{table}
\end{landscape}

% \bibliography{Orbifold}
% \bibliographystyle{NewArXiv}
\end{document}